\documentclass[12pt,preprint]{aastex}
\shorttitle{Precision VLBI pulsar astrometry}
\shortauthors{Deller, Tingay, \& Brisken}

\begin{document}

\newcommand{\pbdot}{\ensuremath{\dot{P}_{\mathrm b}}}
\newcommand{\esys}{\ensuremath{E_{\mathrm{sys}}}}
\newcommand{\etherm}{\ensuremath{E_{\mathrm{therm}}}}
\newcommand{\phisys}{\ensuremath{\phi_{\mathrm{sys}}(t)}}
\newcommand{\degrees}{\ensuremath{^{\circ}}}

\title{Precision Southern Hemisphere pulsar VLBI astrometry: techniques and results for PSR J1559--4438}

\author{A.T. Deller,\altaffilmark{1,2} S.J. Tingay,\altaffilmark{3} \& 
W. Brisken\altaffilmark{4}}
\altaffiltext{1}{Centre for Astrophysics and Supercomputing, Swinburne University of Technology, Mail H39, P.O. Box 218, Hawthorn, VIC 3122, Australia}

\altaffiltext{2}{co-supervised through the Australia Telescope National Facility, P.O. Box 76, Epping, NSW 1710, Australia}

\altaffiltext{3}{Department of Imaging and Applied Physics, Curtin University of Technology, Bentley, WA, Australia}

\altaffiltext{4}{National Radio Astronomy Observatory, Charlottesville, VA, USA}

\begin{abstract}
We describe a data reduction pipeline for VLBI astrometric observations of pulsars, implemented using
the ParselTongue AIPS interface.  The pipeline performs calibration (including ionosphere modeling), phase referencing with proper accounting of reference source structure, amplitude corrections for pulsar scintillation, and position fitting to yield the position, proper motion and parallax.  The optimal
data weighting scheme to minimize the total error budget of a parallax fit, and how this
scheme varies with pulsar parameters such as flux density, is also investigated.  
The robustness of the techniques employed are demonstrated with the presentation of the first 
results from a two year astrometry program using the Australian Long Baseline Array (LBA).
The parallax of PSR J1559--4438 is determined to be $\pi = 0.384 \pm 0.081$\ mas $(1\sigma)$, 
resulting in a distance estimate of 2600 pc which is consistent with earlier DM and HI absorption 
estimates. 
\end{abstract}

\keywords{Techniques: interferometric --- astrometry --- pulsars: general --- pulsars: individual(J1559--4438)}

\section{Introduction}
Accurate, model independent distances to pulsars are, like many astronomical distance
measurements, highly prized but difficult to obtain.  The dispersion measure (DM) of 
a pulsar indicates the integrated column density of free electrons between the observer
and the pulsar, and can be used in conjunction with Galactic electron distribution models
to estimate pulsar distances, but this approach suffers from considerable uncertainty
when the electron distribution model is poorly understood, such as at high Galactic latitudes.
Timing residuals for millisecond pulsars (MSPs) can be used to determine parallax and
proper motion \citep[e.g.][]{hot06}, but only for the very limited subset of pulsars with extremely accurate
timing solutions.  VLBI astrometry offers a means to directly measure the parallaxes and proper
motions of nearby pulsars.

By obtaining an independent pulsar distance estimate, a pulsar DM can be used to determine 
the average electron density along the line of sight; an ensemble of such measurements can be 
used to improve electron distribution models. Whilst several recent large pulsar parallax programs
have significantly increased the number of known VLBI parallaxes \citep[e.g.][]{cha04}, to date 
very few VLBI parallaxes have been obtained for southern pulsars, due to the bias of VLBI facilities 
towards the Northern Hemisphere.  As such, models of Galactic electron distributions are more 
uncertain at far southern declinations.  

Furthermore, binary pulsars offer the opportunity to make exceedingly precise tests of gravitational theories,
since relativistic effects contribute to the observed rate of change of binary period (\pbdot), which
can be measured very precisely in systems with a stable pulsar.  However, 
kinematic effects \citep{shk70} also contribute to \pbdot, and can only
be accurately subtracted to obtain \pbdot\ due to General Relativity (GR) if 
the pulsar distance and transverse
velocity are accurately known.  The effect scales with the square of proper motion, and hence
is typically largest for nearby pulsars, where VLBI measurements of parallax are most feasible.
Since the uncertainty of a DM--based distance estimate 
is large for any individual pulsar, a reliable estimate of the error on derived GR quantities 
requires a direct distance determination. Similarly, luminosities for individual pulsars based on 
DM--inferred distances are questionable, so deriving accurate an accurate luminosity for any 
individual pulsar generally requires an independent confirmation of distance.

In this paper, we describe our target selection policy in \S\ref{sec:select}, and the observations
undertaken in \S\ref{sec:obs}.  The data reduction pipeline is described in \S\ref{sec:datared},
and we present the results of PSR J1559--4438 in \S\ref{sec:results}.  We analyse the optimum
visibility weighting scheme to use for VLBI astrometry in \S\ref{sec:optweight}, and 
the magnitudes of different sources of systematic errors are investigated in \S\ref{sec:errorbudget}.
Our conclusions are presented in \S\ref{sec:conclusions}.

\section{Target selection}
\label{sec:select}
Our observational program encompassed 8 pulsars, listed in Table~\ref{tab:targets}.  
Previous Southern Hemisphere VLBI pulsar astrometry programs
\citep{dod03,leg02,bai90} have yielded only two published pulsar parallaxes, so one of our
primary motivations was to increase the number of southern VLBI parallaxes and hence
improve models of Galactic electron distributions at southern declinations.  Additionally,
as this is the first such large scale southern parallax study, we chose to target pulsars
of varying brightness and predicted distance to determine the types of targets 
that would be feasible for future southern VLBI studies.

In this paper we present results for the bright pulsar J1559--4438.  As an isolated pulsar
with a characteristic age of 4 Myr and DM distance of 2.35 kpc, J1559--438 is unremarkable, but as
it provides the highest signal to noise detections of our target pulsars, and 
suffers most heavily from scintillation, it provides the clearest example of the techniques
which we will apply to the data reduction of the remaining pulsars.  PSR J2048--1616 was
included as a second ``technique check" source and those results will be presented in a future publication.

The remaining targets can be broadly divided into two scientific categories: 
binary pulsars used for tests of GR and gravitational wave detection 
(J0437--4715, J0737--3039A/B and J2145--0750), and pulsars with unusual DM--based luminosity
in the radio (low luminosity pulsars J0108--1431 and J2144--3933) or x-ray 
(J0630--3834, whose x--ray luminosity is anomalously high).
Each group will be the subject of a forthcoming analysis. 

\section{Observations}
\label{sec:obs}
Eight observational epochs were spread over a two year period between May 2006 and February 2008.
Epochs were typically 24 hours in duration, and observed
a subset of the 8 pulsars, depending on which were closest to parallax extrema.
The observing frequency was centred on 1400 MHz for the first observation, and 1650 MHz 
for the remaining 7 observations.   PSR J0437--4715 was observed separately, with four epochs of
12 hours duration, centred on 8400 MHz.  Observations at this higher frequency were made possible
by the high flux and narrow pulse profile of PSR J0437--4715 .  Dual circular polarization was used at all
frequencies.

The Australian Long Baseline Array (LBA) consists of six antennas -- the 
Australia Telescope National Facility (ATNF) telescopes in New South Wales (Parkes, Australia Telescope Compact Array [ATCA], Mopra); the University of 
Tasmania telescopes at Hobart, Tasmania and Ceduna, South Australia; and the
NASA DSN facility at Tidbinbilla, Australian Capital Territory.
The Parkes, phased ATCA, Mopra, and Hobart telescopes participated in all experiments, 
but a Tidbinbilla antenna (70m or 34m) was used only when available, and Ceduna 
participated in observations of J0437-4715 only\footnote{Ceduna 
does not possess a 1600 MHz receiver, so could only participate in the higher frequency
experiments}.  The maximum baseline length with Ceduna is 1700 km, and without Ceduna 
is 1400 km.  Representative $uv$\ coverage at 1650 MHz and 8400 MHz is shown below in 
Figure~\ref{fig:uvcoverage}.

All observations used the recently introduced LBADR disk-based recording system
(Phillips et al., in preparation).
At the three ATNF observatories, the presence of two Data Acquisition System (DAS) units allowed
a recording rate of 512 Mbps (8 $\times$ 16 MHz bands, Nyquist sampled at 2 bits), while the
non-ATNF stations recorded at 256 Mbps (4 $\times$ 16 MHz bands).  For epochs 
where dual--polarization
feeds were available at all antennas, two frequency bands were dropped at the non--ATNF
stations, but as some of the NASA DSN feeds are single polarization only, 1650 MHz 
epochs featuring the 70m NASA antenna and 8400 MHz epochs featuring the 34m NASA
antenna instead retained a single polarization of all frequency bands.  

A phase reference 
cycle time of six minutes, apportioned equally to target and calibrator, was used for all observations.
As the LBA consists of disparate antennas ranging up to 70m in diameter (and, when phased, the 
ATCA has an equivalent diameter of hundreds of metres for the purposes of calculating 
field of view), it was not possible to utilise in--beam calibrators for any sources, unlike
recent pulsar astrometric programs using the VLBA \citep{chatterjee01a,chatterjee05a}.

The data were correlated using matched filtering on pulse profiles with the DiFX 
software correlator \citep{del07}, producing 
RPFITS\footnote{http://www.atnf.csiro.au/computing/software/rpfits.html} format
visibility data.  
Matched pulse profile filters (a more advanced pulsar `gate') 
allow the maximum recovery
of signal to noise when observing pulsars, by dividing the pulse into bins and appropriately
weighting each bin by the expected signal strength before summation.  Table~\ref{tab:targets}
shows the predicted gain due to pulse profile filtering for each target source.  Pulsar ephemerides
were obtained using the ATNF Pulsar Catalogue \citep{man05}, with the exception of the double
pulsar J0737-3039, which was periodically updated with the latest published ephemeris. 
Two second integrations and 64 spectral points per 16 MHz band were used for all observations.

\section{Data reduction}
\label{sec:datared}
Data reduction was performed principally in AIPS\footnote{http://www.aoc.nrao.edu/aips}, 
using the python interface ParselTongue \citep{ket06}.  The DIFMAP package \citep{shep97}
was used for imaging and self calibration.  The data reduction was implemented
as a pipeline, with user interaction for imaging, editing of solution tables, and visibility flagging.  
All scripts used in the data reduction process are available
at \verb+http://astronomy.swin.edu.au/~adeller/software/scripts/+. 
The individual stages of the pipeline are described below.

\subsection{Amplitude and weight calibration and flagging}
Amplitude calibration using the measured telescope system temperatures was carried out using
the AIPS tasks APCAL and ANTAB.  Flagging based on predicted or (when available) logged
telescope off-source times due to slewing or failures was applied using UVFLG, while the first
and last 10 seconds of every scan was excised with the task QUACK.  For the ATCA,
the tied array infrastructure required flagging the 
first 3 correlator integrations (30 seconds) of each scan with QUACK. The weight of each 
visibility point, which is set to a constant value when the RPFITS format data is loaded into
AIPS, was initially scaled by the predicted baseline sensitivity using a ParselTongue script.  The effect of
data weighting is investigated further in \S\ref{sec:dataweights}.

\subsection{Geometric model and ionospheric corrections}
\label{subsec:geoionocorrect}
At the low frequencies which are generally used for pulsar astrometry, ionospheric variations
usually make the dominant contribution to systematic error \citep[see e.g.][]{bri02}.  Using
ionospheric models based on Global Position System (GPS) data 
provided by the NASA Jet Propulsion Laboratory\footnote{available 
from the Crustal Dynamics Data Information Systems (CDDIS) archive: 
ftp://cddis.gsfc.nasa.gov/pub/gps/products/ionex/}, 
we corrected for phase variations due to the ionosphere using the task TECOR.
This observing program roughly coincided with the solar minimum of 2006,
and consequently the ionospheric variations were generally at a minimum.

Total Electron Content (TEC) models suffer at southern declinations due to the relatively
sparse distribution of GPS receivers at southern latitudes.  Consequently, the 
derived ionospheric corrections are much less reliable for the LBA than for similar 
Northern Hemisphere instruments.  The dispersive delay corrections generated 
by TECOR were inspected for each epoch, and the effectiveness
of different TEC maps is investigated in \S\ref{1559:tecor}.

While the observational program was underway, considerably more accurate station
positions were derived for several LBA antennas using archival geodetic observations and the 
OCCAM software \citep{tit04}, a dedicated 22 GHz LBA geodetic experiment (Petrov et al., in
preparation) 
and GPS measurements.  Additionally, more accurate positions for some calibrators were 
published in the 5th VLBA Calibrator Survey \citep[VCS5;][]{kov07}.  The visibilities
were corrected to account for the revised positions using an AIPS SN table generated
with the Wizardry feature of ParselTongue, which stored the difference between the initial
and corrected geometric models.

For some pulsars, their proper motions ($> 100$ mas yr$^{-1}$)\ caused significant position shifts
over the course of a 24 hour observation, comparable in some cases to the epoch positional
accuracy.  As the geometric model generation used in DiFX
at the time of these observations could not account for proper motion, the visibility phases and
uvw values were corrected in AIPS using a SN table generated by ParselTongue, interpolating
between predicted postions for the pulsar at the start and end of the experiment.

\subsection{Fringe-fitting and amplitude calibration refinement}
Fringe fitting was performed using the AIPS task FRING, using a point source model, on the phase reference calibrator data for each target pulsar.
Subsequently, a single structural model was produced for each calibrator using the combined 
datasets from all epochs.  Each source was modeled using a dominant component (delta or narrow
Gaussian) fixed at the phase center, and 0 -- 2 secondary components which were allowed to vary
in position.  The flux of all components was allowed to vary.  Typically, the variation between epochs
of secondary component(s) flux were $<1\%$ of peak image flux.
The solutions were applied to each calibrator and the data averaged in frequency,
exported to disk and loaded into  DIFMAP.  The calibrator model was loaded and 
several iterations of self-calibration and modelfitting performed.  The difmap `modelfit'
command uses the Levenberg--Marquardt least--squares minimisation algorithm to fit the
free model parameters to the visibility points, incorporating the visibility weights.  The self calibration 
corrections were then written to disk as an AIPS SN table using the `cordump' patch to 
difmap\footnote{http://astronomy.swin.edu.au/{\char126}elenc/difmap-patches/}.  Additionally,
data points flagged in DIFMAP were collated in a Wizardry script and converted into a 
flag file suitable for the AIPS task UVFLG.

The amplitude corrections generated in this manner allowed compensation for 
imperfectly measured system temperature values, and were also applied to visibility weights.  
For bands which could not be self--calibrated, due to only being observed by the three 
ATNF antennas, a correction was estimated based on the nearest band with self
calibration solutions of the same polarization.  The self calibration solutions were 
loaded into AIPS using TBIN, and applied to the target pulsars using CLCAL.

The use of bandpass calibration was investigated but found to make insignificant difference to
the fitted target position.  The LBA Data Acquisition System (DAS) utilitizes digital filtering and
typical bandpass phase ripple was $<2\,^{\circ}$.  When averaging in frequency, the lowest and
highest 10\% of the band was discarded and the central 80\% of the band averaged with uniform
weight assigned to each channel.

\subsection{Pulsar scintillation correction}
\begin{figure}
\begin{center}
\begin{tabular}{cc}
\includegraphics[width=0.45\textwidth]{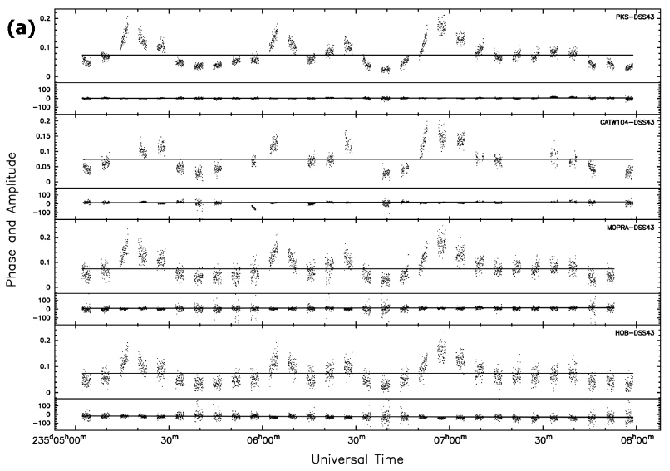} &
\includegraphics[width=0.45\textwidth]{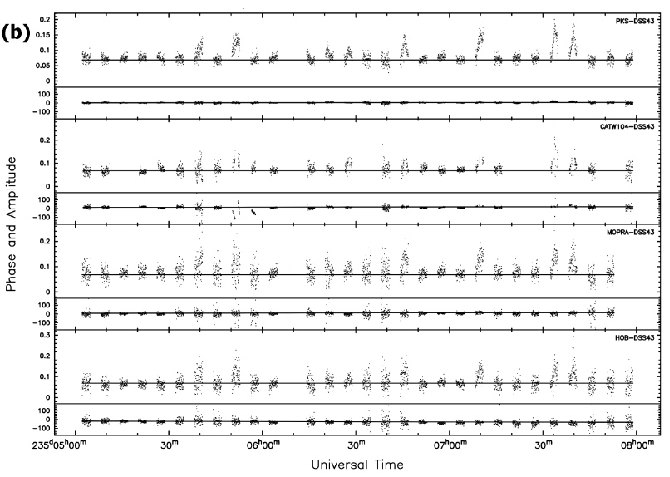} \\
\includegraphics[width=0.45\textwidth]{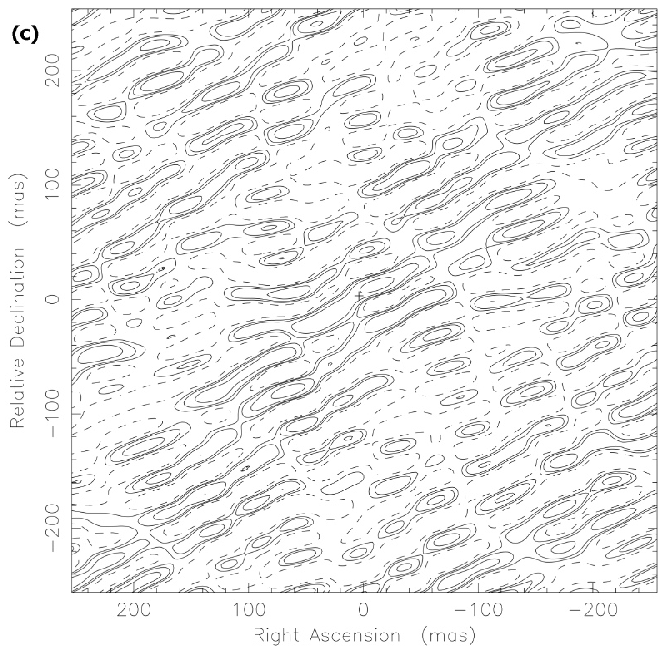} &
\includegraphics[width=0.45\textwidth]{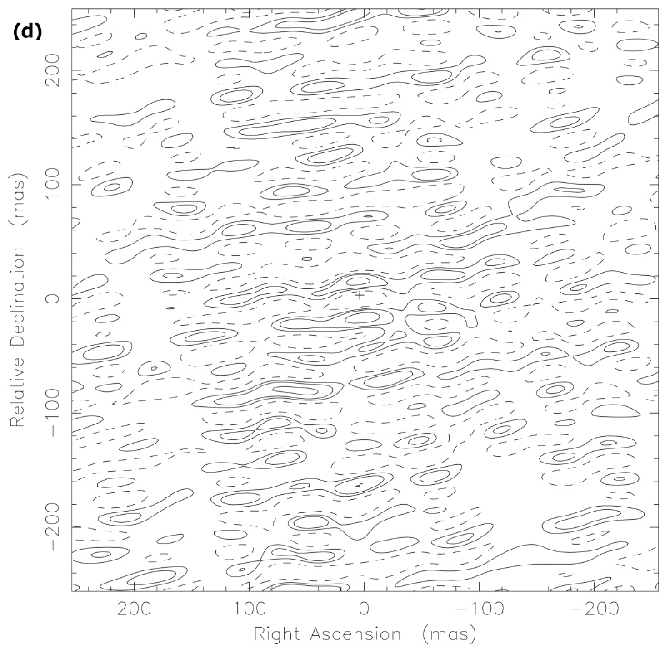} \\
\end{tabular}
\caption{The effects of diffractive scintillation for pulsar J1559--4438.  Panel a) shows the uncorrected
visibility amplitude on baselines including the Tidbinbilla antenna from one experiment -- the 
scintillation timescale of several minutes is apparent.  Panel b) shows the same visibility amplitudes after
correcting for scintillation.  Panel c) shows the image residuals of the uncorrected dataset -- contours are 1,2,4 and 8 mJy/beam.  Panel d) shows image residuals after correcting for scintillation, with contours at 1,2 and 4 mJy/beam -- the improvement in image quality is obvious.}
\label{fig:scint}
\end{center}
\end{figure}

Nearby pulsars can suffer dramatic, and rapid, variations in visibility amplitude due to diffractive
scintillation. The size of the scintillation pattern is typically much larger than the size of the Earth,
and so the amplitude variations are essentially independent of baseline length.  
Maximal amplitude fluctuations (where the rms is equal to the mean
flux) are seen for pulsars in the strong scattering regime \citep[see e.g.][]{walker98a}, which can 
be predicted from Galactic electron distribution models.  The NE2001 model \citep{cordes02a}
predicts that strong scintillation should be observed for J0630--2834, J1559--4438, J2048--1616,
J2144--3933 and J2145--0750.  As an example, Figure~\ref{fig:scint}a shows the variation of 
visibility amplitude with time for J1559--4438 on Tidbinbilla baselines over a 2 hour period.  
Observed scintillation parameters for target pulsars are shown in Table~\ref{tab:scint}.
Assuming the material responsible for the
scintillation is turbulent, with a Kolmogorov distribution, the 
scintillation time $\tau_{\mathrm{scint}}$\ and scintillation bandwidth
$B_{\mathrm{scint}}$\ can be scaled to the frequencies used in these
observations with the relations $\tau_{\mathrm{scint}}\propto\nu^{1.2}$\ 
and $B_{\mathrm{scint}}\propto\nu^{4.4}$\ \citep{cordes86a}.

Left uncorrected, the variation in visibility amplitude with time scatters power throughout the image
plane, as shown in Figure~\ref{fig:scint}c, which shows the residuals for J1559--4438 after fitting a single 
point source to the visibilities shown in Figure~\ref{fig:scint}a.  To overcome this, a ParselTongue
script was written to produce an AIPS SN table which would flatten visibility amplitudes
by averaging data over all sensitive baselines to 1/4 of the scintillation time, normalizing,
and taking the square root to obtain an antenna based correction.  The visibility weights were
then scaled by the inverse of the square of the correction, upweighting points when
the amplitude was high and downweighting points of low significance.  The effect of these 
corrections is shown in Figure~\ref{fig:scint}b and d, which show the visibility amplitudes 
and image residuals respectively.  In this example, the signal to noise ratio (SNR) of the detection is 
improved by a factor of two when the visibility amplitudes are corrected 
for scintillation, which ultimately reduces the error on the position determination 
by a corresponding factor.

\section{Positional determination and parallax fitting}
The calibrated visibilities for each pulsar were averaged in frequency, written to disk
and loaded into DIFMAP.  A single delta function model component was initialized
at the peak of the dirty image, and the modelfit command used to obtain the best fit
for the pulsar visibility data.  Each observing band, as well as the combined dataset, 
was then imaged separately using uniform
weighting (2 pixels, weights raised to the power $-1$) producing 8 images which were saved 
and read back into AIPS using the task FITLD.  
Variations to the weighting scheme are discussed in \S\ref{sec:dataweights}.
The AIPS task JMFIT was used to determine the pulsar position and formal 
(SNR--based) errors in the image plane.  
Systematic offsets of several (3--6) mas were observed at all epochs 
between the 4 bands that were only contributed to by the three ATNF 
antennas, and the 4 bands common to all antennas.  The magnitude and 
direction of these offsets varied between epochs, and so the ATNF--only 
bands (which additionally had formal errors of approximately five times 
the other bands, primarily due to the shorter baselines) were dropped.   
We suspect that the systematic offsets were caused by the lack of an accurate amplitude 
self--calibration solution for these bands.

For each pulsar, the best fit values of J2000 position (RA and declination), proper motion 
(RA and declination) and parallax were initially determined by iteratively minimising the error
function calculated by summing the value of predicted minus actual position
over all measurements, weighted by the individual measurement errors.  The iterative
minimisation code used is described in \citet{bri02}. A reference time
for the proper motion was chosen to be 31 Dec 2006 (MJD 54100) to minimize
proper motion uncertainty contibutions to be position uncertainty.

This approach yields error estimates for the 5 fitted parameters which are almost certainly
an underestimate, for two reasons:
\begin{enumerate}
\item There are systematic errors, varying from band to band within an epoch (intra--epoch
errors) such as the residual unmodeled differential ionosphere, bandpass effects etc, and 
varying from epoch to epoch (inter--epoch errors) caused by effects dependent on 
observing time, such as seasonal or diurnal ionospheric variations, and variations
in refractive scintillation image wander \citep{rickett90a}.  These should increase
the error on each individual measurement, but estimating the magnitude of the systematic
component for each individual measurement is poorly constrained.
\item Each measurement is assumed to be completely independent, whereas as noted above
we expect correlated errors between measurements from the same epoch.  In essence, this
approach overestimates the number of independent measurements, lowering the reduced
chi--squared and implying a better fit than the actual result.
\end{enumerate}
It is possible to make an estimate of the magnitude of the intra--epoch errors by comparing
the scatter in fitted positions from the individual bands.  Forming a weighted centroid
position for each epoch utilising all measurements from that epoch allows an estimation
of the likelihood that the measured points are consistent with that centroid, through the 
calculation of a reduced chi--squared value.  If the reduced chi--squared value exceeds unity,
the presence of unmodeled systematic errors can be inferred.

Since we have no a priori knowledge of the systematic error distribution, we have chosen 
to allocate an equal systematic error to each measurement, and add in
quadrature to the original measurement error. The errors in right ascension and declination 
are treated separately. This is necessarily an iterative procedure, since
the weighted centroid will be altered by the addition of these systematic error estimates.
In effect, this assumes a zero mean, gaussian distribution for the systematic errors.
Although this is unlikely to be the true distribution, it is the most reasonable assumption available, and
certainly more correct than assuming no systematic errors at all.  The intra--epoch systematic
error estimate for an epoch is obtained when the reduced chi--squared reaches unity.

Once a single position measurement and error has been calculated for each epoch,
the fit to position, parallax and proper motion can be re--calculated, and 
the reduced chi--squared of the fit inspected again.   Since the number of 
measurements were reduced, the addition of systematic errors did not always lower
the reduced chi--squared of the fit.  If reduced chi--squared remained significantly
greater than unity, we concluded that significant inter--epoch systematic errors remained.
As with intra--epoch errors, the true error distribution is unknown, and so the inter--epoch
error was apportioned equally between epochs.   The errors in right ascension and declination 
were treated separately, and iterated until a reduced chi--squared of 
unity was obtained.

Thus, the final astrometric dataset for each pulsar consisted of a single position measurement
for each epoch, with a total error equal to the weighted sum of the individual band
formal errors,  added in quadrature to the estimates of intra--epoch and inter--epoch
systematic errors.

For each pulsar, the robustness of error estimation after the inclusion of estimated systematic
contributions was checked by implementing a bootstrap technique.
Bootstrapping, which involves repeated trials on samples selected with replacement 
from the population of measured position points, is a statistical technique allowing the 
estimation of parameter errors without complete knowledge of the underlying distribution
\citep{press02a}.  In this instance, the original single band position measurements
were taken as the population, and N samples were drawn, where N was the original number
of measurements.  Each bootstrap consisted of 10,000 such trials, and the parameter
errors estimated from the variance of the resultant distributions.
A minimum of 3 different epochs needed to be included to ensure a fit was possible - on
the rare occasion that a trial did not satisfy this requirement, it was re--drawn.

Thus, three sets of fitted parameters and errors were obtained for each pulsar -- a ``naive" result
using the single--band positions, a bootstrap result, and a more conservative estimate which
attempts to account for the impact of systematic errors (the ``inclusive" fit).  
In general, the estimated magnitude
of errors on fitted parameters increased through these three different schemes.  
Typically, the ratio in the errors on the inclusive fit to those on the naive fit ranged from 0.95 to 1.90.  
We feel that the final error values obtained from the inclusive fit 
are the most accurate estimation possible, and are inherently conservative.  All quoted 
errors are 1$\sigma$ unless otherwise stated.

\section{Results for PSR J1559--4438}
\label{sec:results}

\subsection{Initial results}
\label{sec:initresults}
\begin{figure}
\begin{center}
\includegraphics[width=0.8\textwidth]{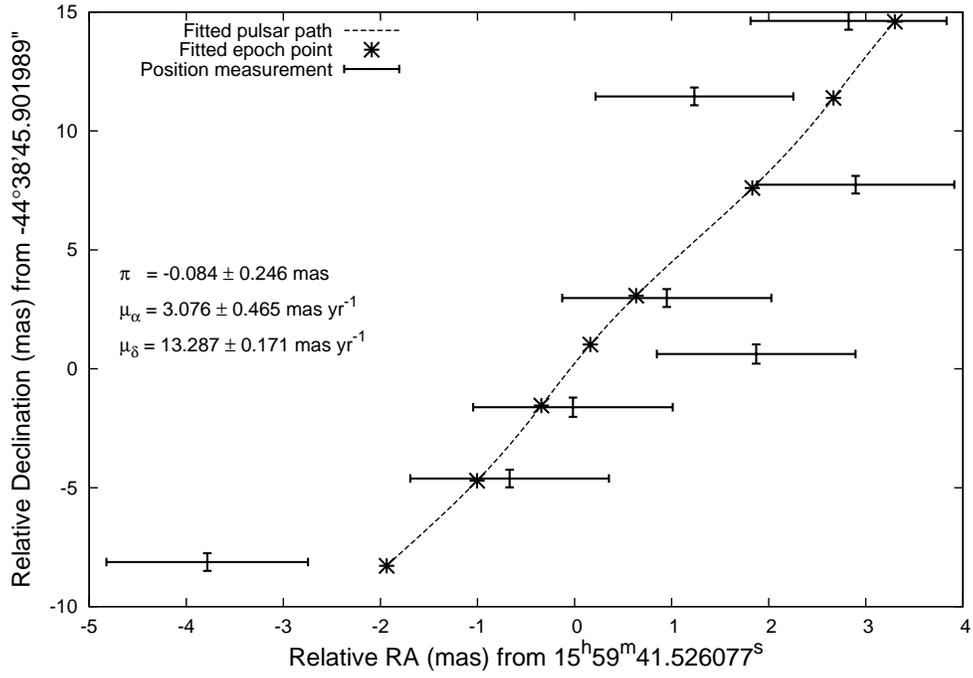}
\caption{Motion of the pulsar in right ascension and declination, with measured positions
overlaid on the best fit.  Sensitivity--weighted visibilities were used. 
The motion of the pulsar is positive in right ascension and declination.
The first epoch (lower left) is clearly inconsistent.}
\label{fig:weightedfit}
\end{center}
\end{figure}

Using the techniques described above, we obtained initial results for J1559--4438 
(shown in Table~\ref{tab:initial}).  The motion of J1559--4438
in right ascension and declination is shown in Figure~\ref{fig:weightedfit}, along with the fitted path
according to the systematic--error weighted fit.
The fit is clearly unsatisfactory, since it predicts a negative parallax (though consistent with zero).
The final column of Table~\ref{tab:initial} shows that systematic errors far exceed the nominal
single--epoch positional accuracies, and inspection of Figure~\ref{fig:weightedfit} shows that 
the first epoch (MJD 53870) is markedly discrepant with the remaining epochs.

As we show below, fine-tuning of the data reduction is required in order to obtain optimal results from each pulsar, in particular with regard to the details of the ionospheric corrections and the visibility weighting schemes using in imaging.  We describe the steps taken for J1559--4438 in 
\S\ref{1559:tecor} and \S\ref{sec:dataweights}.

\subsection{Ionospheric correction}
\label{1559:tecor}

The obvious source of the large systematic errors present in the initial fit shown in 
\S\ref{sec:initresults} is the varying ionospheric correction between epochs.  Accordingly,
as a first check, the position fits for each epoch were recalculated after subtracting the differential
ionospheric correction between PSR J1559--4438 and its phase reference -- in effect,
removing the applied ionospheric correction and leaving the data uncorrected for
ionospheric effects.  This was implemented
using a ParselTongue script which made a two--point interpolation between adjacent calibrator scans
to calculate the differential correction to the target (which was not absorbed into the fringe--fit),
which was stored in a CL table and subtracted using the AIPS task SPLAT.  The applied values
were saved for later analysis.

\begin{figure}
\begin{center}
\includegraphics[width=0.8\textwidth]{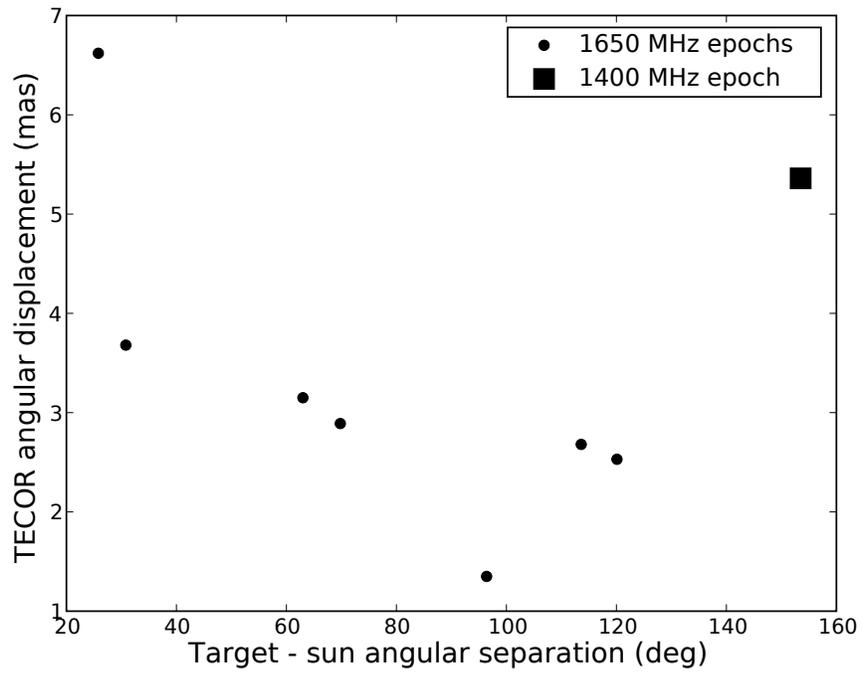}
\caption{Variation of fitted position shifts due to ionospheric correction for PSR J1559--4438 vs 
angular separation between the pulsar and the Sun.  The single 1400 MHz epoch has an
unusually large correction despite the large angular separation between the pulsar and the Sun.}
\label{fig:tecor_sun}
\end{center}
\end{figure}

\begin{figure}
\begin{center}
\includegraphics[width=0.8\textwidth]{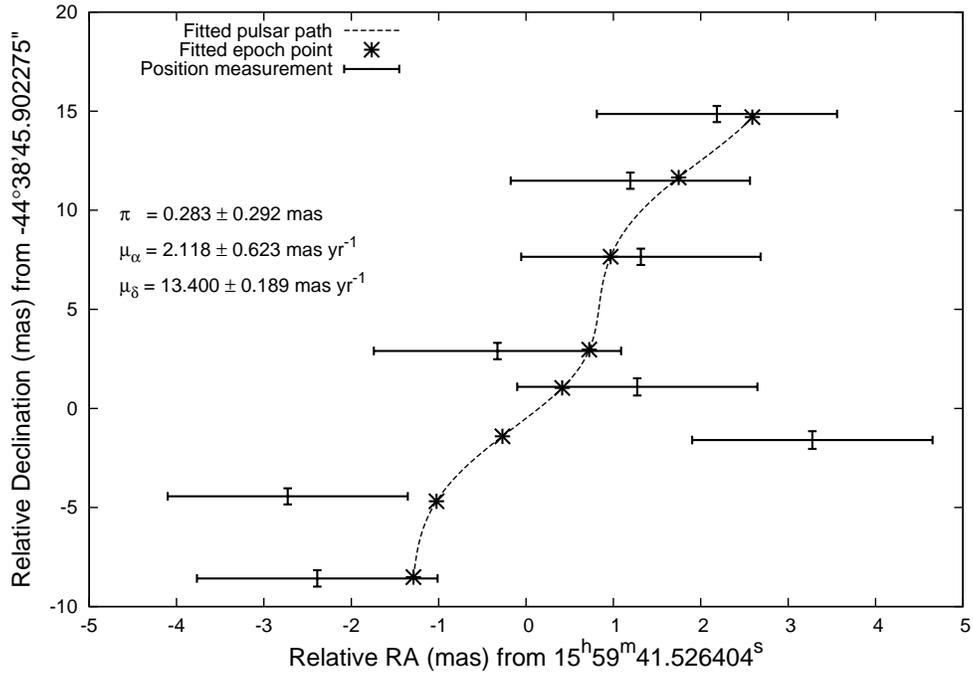}
\caption{Motion of the pulsar in right ascension and declination, with measured positions
overlaid on the best fit, when ionospheric corrections have been removed.  
Sensitivity--weighted visibilities were used. 
The first epoch (lower left) is now more consistent, but the third epoch (during which the pulsar--Sun
angular separation was only $26\,^{\circ}$) is now inconsistent.}
\label{fig:notecorfit}
\end{center}
\end{figure}

Typically, we would expect the largest ionospheric corrections when the angular displacement of
the pulsar from the sun is small, since angular displacement and solar activity are 
the largest influence on TEC.  The angular displacement at each epoch between the original fitted 
position and the position obtained when ionospheric correction was 
removed is presented in Table~\ref{tab:tecorshifts}, and
plotted against angular separation of the pulsar from the sun at the time of observation in
Figure~\ref{fig:tecor_sun}.  The revised astrometric fit obtained without ionospheric correction 
is plotted in Figure~\ref{fig:notecorfit}.

It is immediately apparent from Figure~\ref{fig:tecor_sun} that the first epoch 
(MJD 53870) is discrepant in that
the position shift due to ionospheric correction is unusually large, given the large angular
separation from the Sun.  As shown in Figure~\ref{fig:notecorfit}, this epoch becomes more 
consistent with the fit when the ionospheric correction is removed, 
but the third epoch (MJD 54057) becomes much more
inconsistent.  This is unsurprising, however, since this epoch had the smallest pulsar--Sun separation
and the largest ionospheric corrections.  

To investigate whether the chosen ionospheric map was at fault, the first epoch was re--reduced with
all available maps from the NASA CDDIS 
archive\footnote{ftp://cddis.gsfc.nasa.gov/pub/gps/products/ionex/}, but no significant change 
was found in fitted position.  Given that the different TEC maps make use of many of the same
GPS stations, this is unsurprising.  This problem is exacerbated at southern declinations due
to the low density of GPS receivers at southern latitudes.  
Additionally, any errors in the TEC maps would have an
impact $\sim40\%$ greater for this first epoch, due to its lower observing frequency of 1400 MHz,
compared with the 1650 MHz center frequency used for all subsequent observations.

\begin{figure}
\begin{center}
\includegraphics[width=0.8\textwidth]{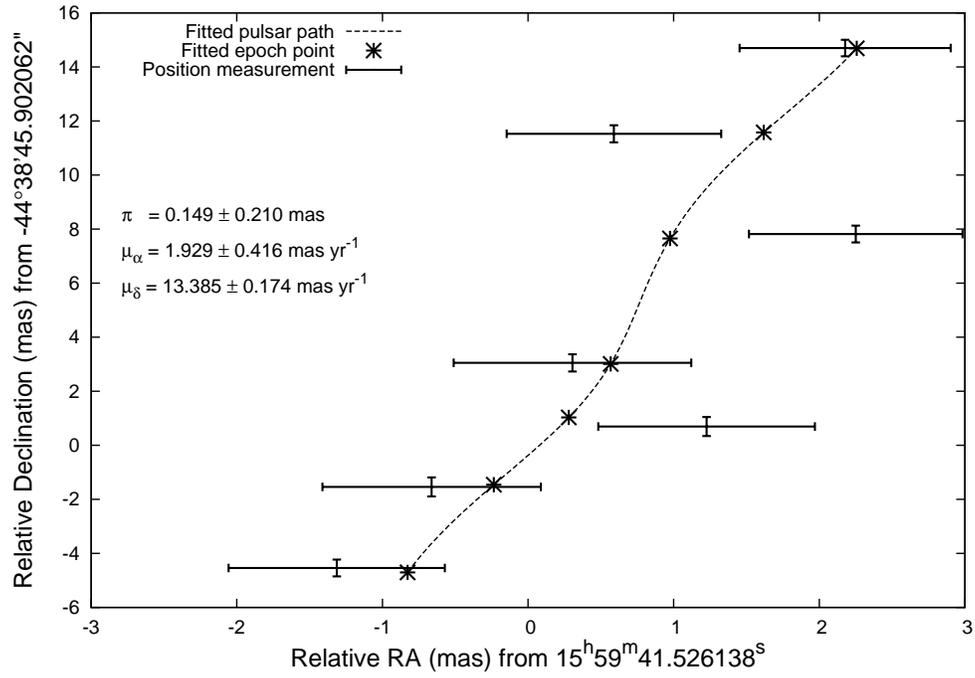}
\caption{Motion of the pulsar in right ascension and declination, with measured positions
overlaid on the best fit, with ionospheric corrections reinstated but the first epoch dropped.  
Sensitivity--weighted visibilities were used. 
The fit is improved considerably.}
\label{fig:weighted_no_v190b}
\end{center}
\end{figure}

Thus, due to the probability of residual ionospheric errors for this epoch, and also
the potential for frequency--dependent calibrator source structure, the first epoch 
(MJD 53870, the only 1400 MHz epoch) was dropped from all further analysis.  
The fit to the remaining 7 epochs, with
ionospheric corrections re--enabled, is shown in Figure~\ref{fig:weighted_no_v190b}.  While
a realistic fit is now obtained, the measurement of parallax is still not significant.

\subsection{Data weights}
\label{sec:dataweights}

Initial pulsar imaging and position fitting used visibilities weighted according
to the best estimate of instantaneous baseline sensitivity.  Whilst this is theoretically optimal 
for data which consists only of signal $S$ and additive random thermal errors \etherm, 
it fails to account for the presence of multiplicative 
systematic errors $\esys=e^{i \phisys}$ caused by 
residual calibration errors.  Typically, these systematic
errors are dominated by atmospheric and ionospheric gradients, although other contributions
include antenna and calibrator source position errors, time--variation of antenna bandpasses, 
and instrumental phase jitter. \citet{fomalont05a}
presents a theoretical review of phase referencing errors, while \citet{pradel06a} presents a 
simulation--based approach.

If \phisys\ had zero--mean and was ergodic, its effect would be
indistinguishable from thermal noise and could be easily estimated, allowing the visibility weights
to be corrected.  For antenna/source position errors
and large--scale atmospheric/ionospheric structure, however, the residual errors are correlated
over long times, causing systematic shifts in the fitted position for the target.

When normal sensitivity--based 
weighting is employed in the presence of substantial and persistent systematic
phase errors, the systematic noise on the most sensitive baseline will be absorbed into the fit, at the cost 
of a poorer fit to the less sensitive baselines.  The magnitude of the induced error will be
dependent on the ratio of systematic to thermal errors, and the discrepancy in sensitivity
between the most and least sensitive baselines in the array.   For the LBA,
the most sensitive baseline (Parkes--DSS43: system equivalent flux density 30 Jy) 
is roughly 13 times more sensitive than 
the least sensitive (Hobart--Mopra: 380 Jy).  Thus, the LBA is particularly susceptible to 
the influence of systematic errors, due to the pronounced variation in baseline sensitivities.

\begin{figure}
\begin{center}
\includegraphics[width=0.8\textwidth]{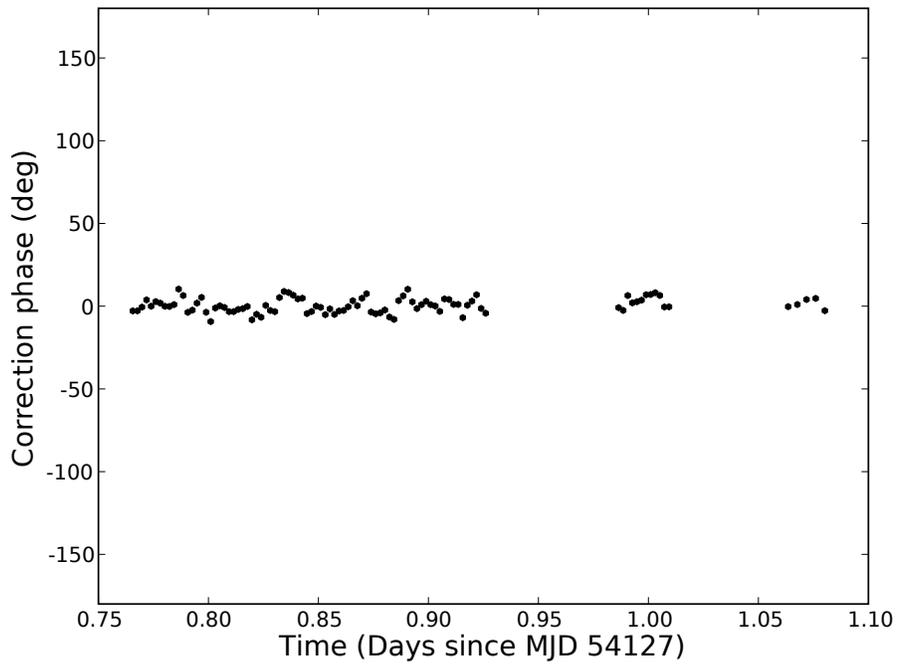}
\caption{Self--calibration corrections, using a three--minute timescale, 
for the ATCA station on J1559--4438.  Clear systematic 
deviations are seen from the zero--mean distribution expected from purely thermal noise.
The rms of the corrections exceeds those expected due to thermal noise by an order of magnitude.}
\label{fig:selfcal}
\end{center}
\end{figure}

The systematic errors can be crudely estimated (in a model--dependent fashion) 
by performing phase--only self--calibration on the target pulsar over a sufficiently 
long timescale to obtain sufficient SNR,
and comparing the magnitude of the corrections to those expected from thermal noise alone.
While this approach probes systematic
errors over a shorter time period than those which would dominate for inter--epoch errors
(tens of minutes, rather than hours to days), it is illustrative of the presence of systematic 
errors overall.
Such corrections are shown in Figure~\ref{fig:selfcal} for the ATCA station using a three--minute
solution interval during the observation on MJD 54127.  The corrections are clearly correlated 
over timescales of tens of minutes, and
the rms deviation of $4.4\,^{\circ}$\ is an order of magnitude greater than the estimated thermal 
phase rms of $0.4\,^{\circ}$\ (calculated in this high--SNR limit as station sensitivity divided
divided by target flux, scaled by pulse filtering gain and converted from radians to degrees).  
Thus, for this observation, systematic errors $\gg$ thermal errors and weighting visibilities 
by sensitivity actually degrades the quality of the position fit.

If the average \esys\ could be accurately estimated for each baseline over the duration of an 
experiment, the baseline visibility weights could be adjusted by assuming that \esys\ is time--invariant.
Even more desirable would be the estimation of \esys\ as a function of time, allowing 
a time--variable adjustment of the visibility weights.
Given present instrumentation, there is no way to reliably estimate systematic 
error (time dependent or independent) in a model independent fashion.
In the limit where $\esys\gg\etherm$, however,  
the visibility weight for each baseline (regardless of sensitivity) 
will be dominated by the systematic error contribution, resulting in
approximately equal weights for all visibilities.  Accordingly, visibility weights for all
baselines were reset to an equal, constant value and data reduction repeated.  
The results are discussed below in \S\ref{1559:final}.

\subsection{Final results}
\label{1559:final}

\begin{figure}
\begin{center}
\includegraphics[width=0.8\textwidth]{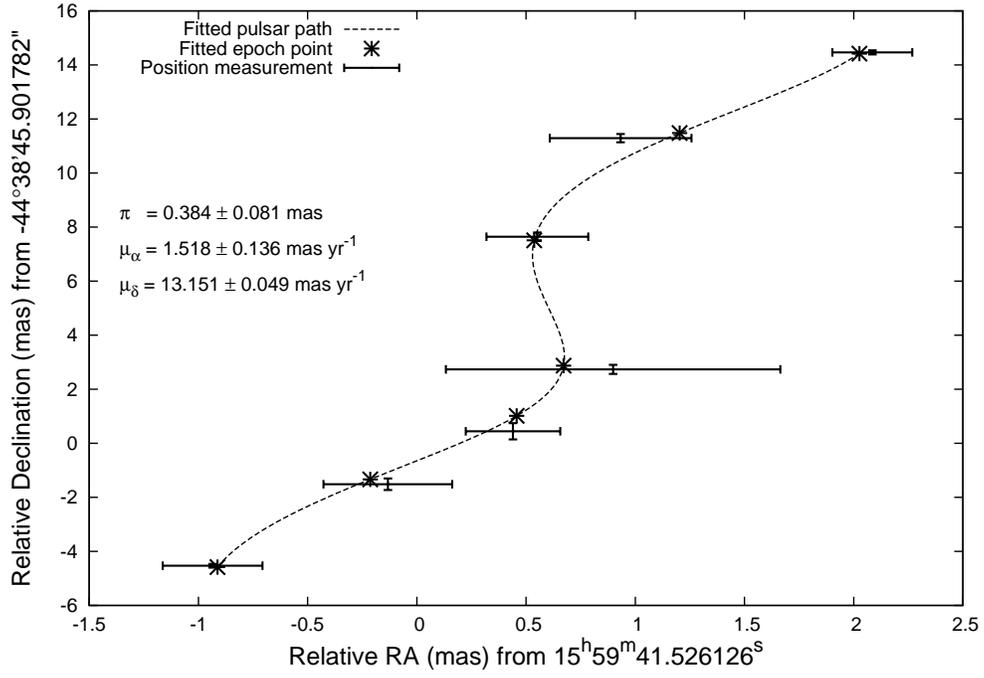}
\caption{Motion of the pulsar in right ascension and declination, with measured positions
overlaid on the best fit, with ionospheric corrections reinstated but the first epoch dropped.  
Equally weighted visibilities were used, mitigating systematic errors and allowing for the first
time a significant measurement of the parallax for this pulsar.}
\label{fig:final}
\end{center}
\end{figure}

The revised fit obtained using equally weighted visibility data is described in Table~\ref{tab:final} and
plotted in Figure~\ref{fig:final}.  Through comparison of Table~\ref{tab:final} with Table~\ref{tab:initial},
it can be seen that the average fit error for a single epoch has increased by 20\%, but the 
inter--epoch systematic error has decreased by 95\%.  Thus, while using equally weighted data
incurs a small sensitivity penalty, it benefits significantly through the reduced susceptibility
to systematic errors.

The use of natural weighting, as opposed to uniform weighting, was investigated but found
to produce inferior results.  Fitted parameters remained relatively constant but errors on
the fitted parameters increased by $\sim 50\%$.  This is unsurprising, since the use of
natural weighting promotes a larger beamsize, since more visibility points are 
concentrated at small {\it uv} distances.   

The best--fit distance of $2600^{+690}_{-450}$\ pc is consistent with the DM--based distance
prediction from the NE2001 Galactic electron distribution model 
\citep[2350 pc;][]{cordes02a}, which differed considerably from the earlier 
\citet{tay93} distance estimate of 1580 pc.  Whilst DM--based distance predictions are often assumed
to be accurate to $\sim$20\%, errors up to a factor of several have been observed for
individual objects \citep[e.g. PSR B0656+14;][]{brisken03a}.
The VLBI distance is also consistent with the lower distance estimate of $2.0 \pm 0.5$ kpc made 
using HI line absorptions by \citet{koribalski95a}. The measured values of
proper motion ($\mu_{\alpha} = 1.52 \pm 0.14$\ mas yr$^{-1}$, 
$\mu_{\delta} = 13.15 \pm 0.05$\ mas yr$^{-1}$) 
are consistent with the VLA observations of \cite{fomalont97a}, who measured
$\mu_{\alpha} = 1 \pm 6$\ mas yr$^{-1}$, $\mu_{\delta} = 14 \pm 11$\ mas yr$^{-1}$.
Neglecting acceleration from the Galactic potential, the kinematic age of the pulsar can be estimated 
from its Galactic latitude $b = 6.3667\,^{\circ}$\ and
proper motion perpendicular to the Galactic plane $\mu_{b} = 8.93$\ mas yr$^{-1}$ as 
$2.57 \pm 0.51$ Myr, assuming a birth location within 100 pc of the plane.  
Under the standard assumption of a braking index
of 3, the observed period $P=257$\,ms and period derivative 
$\dot{P} = 1.01916\times10^{-15}$\ \citep{siegman93a} imply a birth period 
$P_{0}$\ between 35 and 151 ms -- longer than is often assumed for normal pulsars
\citep[see e.g.][]{migliazzo02a}, but similar to the calculated value of $P_{0} =139.6$\,ms 
for PSR J0538+2817 \citep{kramer03a}.

With an accurate proper motion now calculated, the position angle of the proper motion
for PSR J1559--4438 can be compared to the position angle of the emission polarisation, which tests
the alignment of the rotation and velocity vector, as described by \citet{johnston07a}.  If
the pulsar emits predominantly parallel or perpendicular to the magnetic field lines,
then the angle between the velocity and polarisation position angles is expected to be
0\degrees\ or 90\degrees\ respectively.  \citet{johnston07a} found plausible alignment
in 7/14 pulsars of similar ages to PSR J1559--4438.  From
Table~\ref{tab:final}, it is easy to calculate the velocity position angle as 
PA$_{v} = 6.6 \pm 0.6$\degrees.  The polarisation position angle for PSR J1559--4438 given in 
\citet{johnston07a} is $71\pm3$, and so in this instance there is no strong case for alignment
of the velocity and rotation axes.

One inconsistency with previously published data on PSR J1559--4438 is the transverse 
velocity, which at 164 km s$^{-1}$ (95\% confidence upper limit of 287 km s$^{-1}$) is 
inconsistent with the 400 km s$^{-1}$ estimated from 
scintillation by \citet{joh98}.  However, this scintillation estimate assumes that the scattering material
resides in a thin screen at the midpoint between the pulsar and the solar system, and neglects 
any motion of the scattering screen itself.  As the true distribution of the 
scattering material along the line of sight to the pulsar is not known, the most reasonable interpretation
of our results is that the scattering screen for PSR J1559--4438
resides considerably closer to the pulsar than to the solar system.

\section{Optimal data weighting}
\label{sec:optweight}
From the results shown in \S\ref{sec:dataweights} and \S\ref{1559:final}, it is clear that 
for PSR J1559--4438 our astrometric error budget is dominated by systematic errors, and that the
use of equally weighted visibility points is optimal.  However, Table~\ref{tab:targets} shows that 
this may not be the case for other pulsars in our target sample, as some are orders of magnitude
fainter than PSR J1559--4438.  Accordingly, we have investigated
the conditions under which sensitivity--weighted visibilities give superior results to
equally--weighted visibilities.  This was carried out by adding simulated
thermal noise of varying RMS to the existing dataset.

Three ``noisy" datasets $D_{A}$, $D_{B}$, and $D_{C}$\ were constructed by adding 
gaussian--distributed noise to the real and imaginary visibility components of the original 
observations.  Since the theoretical single epoch SNR for sensitivity--weighted data should be 
$\sim800$ (a factor of 10 greater than the typically obtained SNR), 
the RMS of the added noise in the three datasets was set to 
predicted baseline sensitivity scaled by a factor of 20, 40, and 80, which should allow
a maximum single--epoch SNR of 40, 20 and 10 respectively.
The results of fitting the modified datasets (using the inclusive fit approach only) with
and without the use of sensitivity weighting are presented in Tables \ref{tab:optweight_with} and
\ref{tab:optweight_without} respectively.

Tables~\ref{tab:optweight_with} and \ref{tab:optweight_without} show that while the equally weighted
dataset performs better for $D_{A}$, when the average epoch SNR is still high, its performance
rapidly deteriorates as the average epoch SNR decreases.  
In $D_{C}$, the pulsar was not detected in several
epochs using equally--weighted data.  The reduction in performance is
less marked for the weighted datasets, although they are clearly still affected by
systematic errors.  However, if the pulsar was closer and the parallax larger, these systematics
would be less dominant, and weighted datasets would allow measurement of a parallax
when equally weighted datasets may be overwhelmed by thermal errors, to the point of not detecting
the pulsar in a single epoch.

As noted in \S\ref{sec:dataweights}, the use of weighted visibility points would always be
optimal if the weights could include an estimate of the systematic error contribution to that
visibility.  In the absence of such an estimate, we propose that for the LBA with typical observing
conditions and calibrator throws, the transition region from systematic to thermal error dominated
astrometry occurs when the single--epoch detection SNR falls to approximately 10 for
equally--weighted visibilities.  This is
shown by the similarity of result quality for $D_{B}$, where the average epoch SNR was approximately 
10 for the equally--weighted visibilities.
Alternatively, both weighting regimes can be used and average total single--epoch error 
(formal + systematic) can be compared to estimate the optimal weighting scheme.
Again, this is borne out in the simulated datasets, where a transition
from systematic errors dominating to epoch fit errors dominating is seen as more noise is added.

\section{Estimated contributions to systematic error}
\label{sec:errorbudget}

The major contributions to systematic error in VLBI astrometry 
include geometric model errors (station/source
position, Earth Orientation Parameters), residual ionospheric and tropospheric errors, 
variable phase reference source structure, and image
wander due to refractive scintillation.  Of these, at 1650 MHz residual ionospheric errors would be
expected to dominate, despite the a priori ionospheric calibration employed 
\citep[e.g.][]{bri02}.
However, these should be largely uncorrelated from epoch to epoch, and hence the addition of
more observations can be expected to continue to improve the fit.  Residual tropospheric errors
should also be uncorrelated with epoch, and considerably smaller.

Geometric model errors cause relative astrometric errors which increase with calibrator throw.  
Earth Orientation Parameters (EOPs) are well determined
by geodetic observations and make minimal contributions to astrometric errors \citep{pradel06a}.
Similarly, the mean position of well-determined calibrators makes a minimal contribution.  
Some LBA stations, however, have position uncertainties of several cm, 
which could make a several hundred $\mu$as 
contribution to systematic error.  The magnitude of the error
depends on the source declination and the calibrator--target separation.  Given similar $uv$\ coverage
at different epochs, however, the offset will be largely constant with time and is absorbed into
the reference position of the target.
Future planned geodetic observations
will continue to improve LBA station positions, and reduce this systematic contribution.  Additionally,
as noted in \S\ref{subsec:geoionocorrect}, small station position errors can be 
corrected post--correlation, which offers the potential to further improve previous position fits.

Refractive image wander is caused by large--scale fluctuations in the ISM, and can
be estimated based on the strength of the pulsar scattering and the scattering disk size
\citep[e.g.][]{rickett90a}.  For strong scattering, which includes PSR J1559--4438, the image
wander is less than the scattering disk size, if Kolmogorov turbulence is assumed for the scattering
material \citep{rickett90a}.  Thus, since Table~\ref{tab:scint} shows that scattering disk of 
PSR J1559--4438 is estimated to be only 133 $\mu$as at our observing frequency, 
the maximum refractive image wander is $\ll 100\ \mu$as, and can be discounted 
as a source of systematic error.  Table~\ref{tab:scint} shows that refractive scintillation is unlikely
to be significant for any of our currently targeted pulsars.

The variability of calibrator structure with time depends on the source chosen, but all compact
extragalactic radio sources are expected to show some variability, with typical
rms values of 100~$\mu$as \citep{fomalont05a}.  Over short time periods, this image wander
may be correlated from epoch to epoch and absorbed into proper motion fits, 
but over long times (which could be longer than
as astrometric observing program), the mean apparent position will be constant.

\section{Conclusions}
\label{sec:conclusions}
A pipeline for the reduction of LBA astrometric data has been developed in ParselTongue and
verified by calculating the parallax ($\pi = 0.384 \pm 0.081$\ mas) and proper 
motion ($\mu_{\alpha} = 1.52 \pm 0.14$\ mas yr$^{-1}$, $\mu_{\delta} = 13.15 \pm 0.05$\ mas yr$^{-1}$)
of PSR J1559--4438.  The calculated values are consistent with the DM distance estimate 
and earlier HI absorption and proper motion studies.  Full account
has been made of the impact of residual systematic errors on the quality of the astrometric
fit.  The optimal weighting scheme in the presence of systematic errors and varying thermal
errors has been investigated, resulting in the guideline that superior astrometric quality can be
obtained for the LBA for typical astrometric observations by using equally weighted, as opposed
to sensitivity weighted, visibilities if the 
target can be detected with S/N $>10$.  The completion of this parallax program 
will result in the publication of 5 more Southern Hemisphere pulsar parallaxes, which
will quadruple the number of Southern Hemisphere pulsars with parallaxes determined directly from VLBI astrometry.

\acknowledgements

The authors would like to thank Ramesh Bhat for useful discussions regarding pulsar scintillation.
This work has been supported by the Australian Federal Government's Major National Research Facilities program.  ATD is supported via a Swinburne University of Technology Chancellor's Research Scholarship and a CSIRO postgraduate scholarship.  The Long Baseline Array is part of the Australia Telescope which is funded by the Commonwealth of Australia for operation as a National Facility managed by CSIRO.  This research made use of the
ATNF Pulsar Catalogue and the NASA/IPAC Extragalactic Database (NED). NED is operated by 
the Jet Propulsion Laboratory, California Institute of Technology, under contract with the 
National Aeronautics and Space Administration.

\clearpage

\begin{deluxetable}{lcccccc}
\tabletypesize{\tiny}
\tablecaption{Target pulsars}
\tablewidth{0pt}
\tablehead{
\colhead{Pulsar} & \colhead{DM distance} & \colhead{1600 MHz} & \colhead{Estimated pulsar} & Equivalent gated & \colhead{Reference} & \colhead{Calibrator/target} \\
\colhead{name} & \colhead{(pc)\tablenotemark{a}} & \colhead{flux (mJy)} & \colhead{gating gain} & 1600 MHz flux (mJy) & \colhead{source} & \colhead{separation (deg)} 
}
\startdata
J0108--1431 		& 130 	& 0.6 \tablenotemark{b} 	& 5.1 	& 2.0		& J0111-1317 	& 1.5 \\
J0437--4715 		& 140 	& 140 				& 6.25	& 875	& J0439-4522  	& 1.9 \\
J0630--2834 		& 2150 	& 23 					& 3.5 	& 81		& J0628-2805  	& 0.7 \\
J0737--3039A/B 	& 570 	& 1.6 				& 2.5 	& 4		& J0738-3025  	& 0.4 \\
J1559--4438 		& 1600 	& 40 					& 3.6 	& 144	& J1604-4441  	& 0.9 \\
J2048--1616 		& 640 	& 13 					& 3.6		& 47		& J2047-1639  	& 0.5 \\
J2144--3933 		& 180 	& 0.8\tablenotemark{b} 	& 10 		& 8		& J2141-3729  	& 2.1 \\
J2145--0750 		& 500 	& 8\tablenotemark{b} 	& 4.3 	& 34		& J2142-0437  	& 3.3 \\
\enddata
\tablenotetext{a}{Taken from Taylor \& Cordes (1993)}
\tablenotetext{b}{Pulsar suffers heavily from long timescale scintillation, so individual epochs vary considerably from the average value shown.}
\label{tab:targets}
\end{deluxetable}

\begin{deluxetable}{lrrrrr}
\tabletypesize{\tiny}
\tablecaption{Observed pulsar scintillation parameters and estimated scattering disk sizes}
\tablewidth{0pt}
\tablehead{
\colhead{Pulsar} & \colhead{Scintillation} & \colhead{Scintillation} & \colhead{Observing} & 
\colhead{Reference} & \colhead{Predicted scattering disk\tablenotemark{a}} \\
\colhead{name} & \colhead{time (s)} & \colhead{bandwidth (MHz)} & \colhead{frequency (MHz)} & 
 & \colhead{size (mas) at 1650 MHz} 
}
\startdata
J0108--1431 	& *\tablenotemark{b}		& *	& *		& Johnston, Nicastro \& Koribalski (1998) & 0.003 \\
J0437--4715 	& 600		& 17.4		& 660 	& Johnston, Nicastro \& Koribalski (1998) & 0.052\\
J0630--2834 	& 456		& 0.2 		& 327	& Bhat, Rao \& Gupta (1999) & 0.022\\
J0737--3039A 	& $>70$\tablenotemark{c}& 1.4\tablenotemark{d} & 1400 	& Coles et al. (2005) & 0.406 \\
J1559--4438 	& 77			& 0.16		& 660	& Johnston, Nicastro \& Koribalski (1998) & 0.133 \\
J2048--1616 	& 138		& 0.54		& 327	& Bhat, Rao \& Gupta (1999) & 0.023 \\
J2144--3933 	& 1500		& 2.9			& 660	& Johnston, Nicastro \& Koribalski (1998) & 0.126 \\
J2145--0750 	& 1510		& 1.48 		& 436	& Johnston, Nicastro \& Koribalski (1998) & 0.046 \\
\enddata
\tablenotetext{a}{Calculated from decorrelation bandwidth where available, and taken from NE2001 model  where scintillation measurements are unavailable}
\tablenotetext{b}{Scintillation parameters have not been measured, but are believed to be extremely large, as expected for a very nearby pulsar}
\tablenotetext{c}{Varies considerable with orbital velocity -- values quoted for highest transverse velocity}
\tablenotetext{d}{Coles, private communication}
\label{tab:scint}
\end{deluxetable}

\begin{deluxetable}{lrrr}
\tabletypesize{\tiny}
\tablecaption{Initial results (sensitivity--weighted visibilities) for PSR J1559--4438}
\tablewidth{0pt}
\tablehead{
\colhead{Parameter} & \colhead{Naive fit} & \colhead{Bootstrap fit} & \colhead{Inclusive fit}
}
\startdata
RA (J2000)	&  15:59:41.526092 $\pm$ 0.000005	
			& 15:59:35.526093 $\pm$ 0.000031	
			& 15:59:41.526077 $\pm$ 0.000026		\\
Dec (J2000) 	&  -44:38:45.902028 $\pm$ 0.000017	
			& -44:38:45.902034 $\pm$ 0.000062	
			& -44:38:45.901989 $\pm$ 0.000102	\\
$\mu_{\alpha}$	(mas yr$^{-1}$)	&  2.64 $\pm$ 0.07\phn\phn\phn\phn	
						& 2.65 $\pm$ 0.44\phn\phn\phn\phn		
						& 3.08 $\pm$ 0.46\phn\phn\phn\phn	   \\
$\mu_{\delta}$	(mas yr$^{-1}$)	&  13.36 $\pm$ 0.02\phn\phn\phn\phn	
						& 13.36 $\pm$ 0.06\phn\phn\phn\phn		
						& 13.29 $\pm$ 0.17\phn\phn\phn\phn	   \\
$\pi$ (mas)	 			&  -0.066 $\pm$ 0.038\phn\phn\phn	
						& -0.054 $\pm$ 0.193\phn\phn\phn	
						& -0.083 $\pm$ 0.245\phn\phn\phn  \\
Nominal distance (pc)		&  -15100 	\phs \phn\phd\phn\phn\phn\phn\phn\phn$\  $			
						& -18500 \phs \phn\phd\phn\phn\phn\phn\phn\phn$\  $		
						& -11900 \phs \phn\phd\phn\phn\phn\phn\phn\phn$\  $	\\
Nominal $v_{t}$ (km s$^{-1}$)	&  -974 \phs \phn\phd\phn\phn\phn\phn\phn\phn$\  $
						& -1190 \phs \phn\phd\phn\phn\phn\phn\phn\phn$\  $
						& -770  \phs \phn\phd\phn\phn\phn\phn\phn\phn$\  $ \\
Reduced chi--squared 		&  8.8 \phs \phn\phd\phn\phn\phn\phn\phn\phn$\  $
						&  \phs \phn\phd\phn\phn\phn\phn\phn\phn$\  $
						& 1.0 \phs \phn\phd\phn\phn\phn\phn\phn\phn$\  $	\\
Average epoch mean fit error(mas)			&  \phs \phn\phd\phn\phn\phn\phn\phn\phn$\  $ 
									&  \phs \phn\phd\phn\phn\phn\phn\phn\phn$\  $
									& 0.198 \phs \phn\phd\phn\phn\phn\phn\phn\phn$\  $ \\
Average intra--epoch systematic error (mas) 	&  \phs \phn\phd\phn\phn\phn\phn\phn\phn$\  $ 
									&  \phs \phn\phd\phn\phn\phn\phn\phn\phn$\  $
									& 0.097 \phs \phn\phd\phn\phn\phn\phn\phn\phn$\  $ \\
Average inter--epoch systematic error (mas) 	&  \phs \phn\phd\phn\phn\phn\phn\phn\phn$\  $ 
									&  \phs \phn\phd\phn\phn\phn\phn\phn\phn$\  $
									& 1.070 \phs \phn\phd\phn\phn\phn\phn\phn\phn$\  $ \\
\enddata
\label{tab:initial}
\end{deluxetable}

\begin{deluxetable}{lccc}
\tabletypesize{\tiny}
\tablecaption{Average position shift due to ionospheric corrections for PSR J1559--4438}
\tablewidth{0pt}
\tablehead{
\colhead{Epoch MJD} & \colhead{RA shift (mas)} & \colhead{Dec shift (mas)} & \colhead{Sun separation (deg)}
}
\startdata
53870	&  $-$5.36	& \phs0.27	& 153.5	\\
53970	&  $-$1.34	& \phs0.15	& \phn96.4	\\
54057	&  $-$6.61	& \phs0.39	& \phn25.8	\\
54127	&  $-$3.15	& $-$0.07		& \phn63.0	\\
54182	&  $-$2.66	& \phs0.35	& 113.6	\\
54307	&  $-$2.51	& \phs0.34	& 120.1	\\
54413	&  $-$3.68	& \phs0.20	& \phn30.8	\\
54500	&  $-$2.89	& \phs0.05	& \phn69.8	\\
\enddata
\label{tab:tecorshifts}
\end{deluxetable}

\begin{deluxetable}{lrrr}
\tabletypesize{\tiny}
\tablecaption{Final results (equally weighted visibilities) for PSR J1559--4438}
\tablewidth{0pt}
\tablehead{
\colhead{Parameter} & \colhead{Naive fit} & \colhead{Bootstrap fit} & \colhead{Inclusive fit}
}
\startdata
RA (J2000)	& 15:59:41.526121 $\pm$ 0.000006	
			& 15:59:35.526121 $\pm$ 0.000008	
			& 15:59:41.526126 $\pm$ 0.000008		\\
Dec (J2000) 	& -44:38:45.901849 $\pm$ 0.000020	
			& -44:38:45.901859 $\pm$ 0.000072	
			& -44:38:45.901778 $\pm$ 0.000035	\\
$\mu_{\alpha}$	(mas yr$^{-1}$)	& 1.62 $\pm$ 0.08\phn\phn\phn\phn	
					& 1.60 $\pm$ 0.16\phn\phn\phn\phn		
					& 1.52 $\pm$ 0.14\phn\phn\phn\phn	   \\
$\mu_{\delta}$	(mas yr$^{-1}$)	&  13.20 $\pm$ 0.02\phn\phn\phn\phn	
					& 13.21 $\pm$ 0.08\phn\phn\phn\phn		
					& 13.15 $\pm$ 0.05\phn\phn\phn\phn	   \\
$\pi$ (mas)	 		&  0.280 $\pm$ 0.048\phn\phn\phn	
					& 0.291 $\pm$ 0.111\phn\phn\phn	
					& 0.384 $\pm$ 0.081\phn\phn\phn  \\
Nominal distance (pc)	& 3570 	\phs \phn\phd\phn\phn\phn\phn\phn\phn$\  $			
					& 3440 \phs \phn\phd\phn\phn\phn\phn\phn\phn$\  $		
					& 2600 \phs \phn\phd\phn\phn\phn\phn\phn\phn$\  $	\\
Nominal $v_{t}$ (km s$^{-1}$)	&  225 \phs \phn\phd\phn\phn\phn\phn\phn\phn$\  $
					& 217 \phs \phn\phd\phn\phn\phn\phn\phn\phn$\  $
					& 164  \phs \phn\phd\phn\phn\phn\phn\phn\phn$\  $ \\
Reduced chi--squared 	&  3.6 \phs \phn\phd\phn\phn\phn\phn\phn\phn$\  $
					&  \phs \phn\phd\phn\phn\phn\phn\phn\phn$\  $
					& 1.0 \phs \phn\phd\phn\phn\phn\phn\phn\phn$\  $	\\
Average epoch mean fit error(mas)			&  \phs \phn\phd\phn\phn\phn\phn\phn\phn$\  $ 
									&  \phs \phn\phd\phn\phn\phn\phn\phn\phn$\  $
									& 0.242 \phs \phn\phd\phn\phn\phn\phn\phn\phn$\  $ \\
Average intra--epoch systematic error (mas) 	&  \phs \phn\phd\phn\phn\phn\phn\phn\phn$\  $ 
									&  \phs \phn\phd\phn\phn\phn\phn\phn\phn$\  $
									& 0.259 \phs \phn\phd\phn\phn\phn\phn\phn\phn$\  $ \\
Average inter--epoch systematic error (mas) 	&  \phs \phn\phd\phn\phn\phn\phn\phn\phn$\  $ 
									&  \phs \phn\phd\phn\phn\phn\phn\phn\phn$\  $
									& 0.055 \phs \phn\phd\phn\phn\phn\phn\phn\phn$\  $ \\
\enddata
\label{tab:final}
\end{deluxetable}

\begin{deluxetable}{lrrr}
\tabletypesize{\tiny}
\tablecaption{Noise--added fits for PSR J1559--4438, sensitivity--weighted visibilities}
\tablewidth{0pt}
\tablehead{
\colhead{Parameter} & \colhead{$D_{A}$} & \colhead{$D_{B}$} & \colhead{$D_{C}$}
}
\startdata
RA (J2000)	& 15:59:41.526163 $\pm$ 0.000027	
			& 15:59:35.526086 $\pm$ 0.000025	
			& 15:59:41.526025 $\pm$ 0.000041		\\
Dec (J2000) 	& -44:38:45.902054 $\pm$ 0.000154	
			& -44:38:45.902131 $\pm$ 0.000163	
			& -44:38:45.901760 $\pm$ 0.000258	\\
$\mu_{\alpha}$	(mas yr$^{-1}$)	& 2.03 $\pm$ 0.46\phn\phn\phn\phn	
					& 1.44 $\pm$ 0.59\phn\phn\phn\phn		
					& 2.83 $\pm$ 0.79\phn\phn\phn\phn	   \\
$\mu_{\delta}$	(mas yr$^{-1}$)	& 13.39 $\pm$ 0.24\phn\phn\phn\phn	
					& 13.40 $\pm$ 0.21\phn\phn\phn\phn		
					& 13.00 $\pm$ 0.44\phn\phn\phn\phn	   \\
$\pi$ (mas)	 		& 0.109 $\pm$ 0.278\phn\phn\phn	
					& 0.270 $\pm$ 0.373\phn\phn\phn	
					& 0.621 $\pm$ 0.579\phn\phn\phn  \\
Nominal distance (pc)	& 9200 \phs \phn\phd\phn\phn\phn\phn\phn\phn$\  $			
					& 3700 \phs \phn\phd\phn\phn\phn\phn\phn\phn$\  $		
					& 1610 \phs \phn\phd\phn\phn\phn\phn\phn\phn$\  $	\\
Nominal $v_{t}$ (km s$^{-1}$)	& 590 \phs \phn\phd\phn\phn\phn\phn\phn\phn$\  $
					& 237 \phs \phn\phd\phn\phn\phn\phn\phn\phn$\  $
					& 102  \phs \phn\phd\phn\phn\phn\phn\phn\phn$\  $ \\
Reduced chi--squared 	& 1.0 \phs \phn\phd\phn\phn\phn\phn\phn\phn$\  $
					& 1.0 \phs \phn\phd\phn\phn\phn\phn\phn\phn$\  $
					& 0.5 \phs \phn\phd\phn\phn\phn\phn\phn\phn$\  $	\\
Average epoch mean fit error(mas)			& 0.441 \phs \phn\phd\phn\phn\phn\phn\phn\phn$\  $ 
									& 0.904 \phs \phn\phd\phn\phn\phn\phn\phn\phn$\  $
									& 1.308 \phs \phn\phd\phn\phn\phn\phn\phn\phn$\  $ \\
Average intra--epoch systematic error (mas) 	& 0.504 \phs \phn\phd\phn\phn\phn\phn\phn\phn$\  $ 
									& 1.412 \phs \phn\phd\phn\phn\phn\phn\phn\phn$\  $
									& 0.863 \phs \phn\phd\phn\phn\phn\phn\phn\phn$\  $ \\
Average inter--epoch systematic error (mas) 	& 0.769 \phs \phn\phd\phn\phn\phn\phn\phn\phn$\  $ 
									& 0.315 \phs \phn\phd\phn\phn\phn\phn\phn\phn$\  $
									& 0.0 \phs \phn\phd\phn\phn\phn\phn\phn\phn$\  $ \\
Average single--epoch SNR 	& 33  \phs \phn\phd\phn\phn\phn\phn\phn\phn$\  $
						& 17  \phs \phn\phd\phn\phn\phn\phn\phn\phn$\  $
						& 9  \phs \phn\phd\phn\phn\phn\phn\phn\phn$\  $\\
\enddata
\label{tab:optweight_with}
\end{deluxetable}

\begin{deluxetable}{lrrr}
\tabletypesize{\tiny}
\tablecaption{Noise--added fits for PSR J1559--4438, equally weighted visibilities}
\tablewidth{0pt}
\tablehead{
\colhead{Parameter} & \colhead{$D_{A}$} & \colhead{$D_{B}$} & \colhead{$D_{C}$}
}
\startdata
RA (J2000)	& 15:59:41.526142 $\pm$ 0.000026
			& 15:59:35.526115 $\pm$ 0.000042	
			& 15:59:41.526268 $\pm$ 0.000084		\\
Dec (J2000) 	& -44:38:45.901745 $\pm$ 0.000076	
			& -44:38:45.902220 $\pm$ 0.000234	
			& -44:38:45.901798 $\pm$ 0.000420	\\
$\mu_{\alpha}$	(mas yr$^{-1}$)	& 0.91 $\pm$ 0.38\phn\phn\phn\phn	
					& 0.22 $\pm$ 0.99\phn\phn\phn\phn		
					& -0.04 $\pm$ 1.20\phn\phn\phn\phn	   \\
$\mu_{\delta}$	(mas yr$^{-1}$)	&  13.18 $\pm$ 0.10\phn\phn\phn\phn	
					& 13.56 $\pm$ 0.32\phn\phn\phn\phn		
					& 12.65 $\pm$ 0.65\phn\phn\phn\phn	   \\
$\pi$ (mas)	 		& 0.544 $\pm$ 0.235\phn\phn\phn	
					& 0.760 $\pm$ 0.597\phn\phn\phn	
					& 2.092 $\pm$ 0.787\phn\phn\phn  \\
Nominal distance (pc)	& 1840 \phs \phn\phd\phn\phn\phn\phn\phn\phn$\  $			
					& 1320 \phs \phn\phd\phn\phn\phn\phn\phn\phn$\  $		
					& 480 \phs \phn\phd\phn\phn\phn\phn\phn\phn$\  $	\\
Nominal $v_{t}$ (km s$^{-1}$)	& 115 \phs \phn\phd\phn\phn\phn\phn\phn\phn$\  $
					& 85 \phs \phn\phd\phn\phn\phn\phn\phn\phn$\  $
					& 29  \phs \phn\phd\phn\phn\phn\phn\phn\phn$\  $ \\
Reduced chi--squared 	& 0.8 \phs \phn\phd\phn\phn\phn\phn\phn\phn$\  $
					& 1.0 \phs \phn\phd\phn\phn\phn\phn\phn\phn$\  $
					& 0.9 \phs \phn\phd\phn\phn\phn\phn\phn\phn$\  $	\\
Average epoch mean fit error(mas)			& 0.618 \phs \phn\phd\phn\phn\phn\phn\phn\phn$\  $ 
									& 0.937 \phs \phn\phd\phn\phn\phn\phn\phn\phn$\  $
									& 1.853 \phs \phn\phd\phn\phn\phn\phn\phn\phn$\  $ \\
Average intra--epoch systematic error (mas) 	& 0.791 \phs \phn\phd\phn\phn\phn\phn\phn\phn$\  $ 
									& 1.433 \phs \phn\phd\phn\phn\phn\phn\phn\phn$\  $
									& 1.531 \phs \phn\phd\phn\phn\phn\phn\phn\phn$\  $ \\
Average inter--epoch systematic error (mas) 	& 0.000 \phs \phn\phd\phn\phn\phn\phn\phn\phn$\  $ 
									& 0.380 \phs \phn\phd\phn\phn\phn\phn\phn\phn$\  $
									& 0.000 \phs \phn\phd\phn\phn\phn\phn\phn\phn$\  $ \\
Average single--epoch SNR 	& 20  \phs \phn\phd\phn\phn\phn\phn\phn\phn$\  $
						& 10  \phs \phn\phd\phn\phn\phn\phn\phn\phn$\  $
						& 5  \phs \phn\phd\phn\phn\phn\phn\phn\phn$\  $ \\
\enddata
\label{tab:optweight_without}
\end{deluxetable}

\clearpage
\begin{figure}
\begin{center}
\begin{tabular}{cc}
\includegraphics[width=0.4\textwidth]{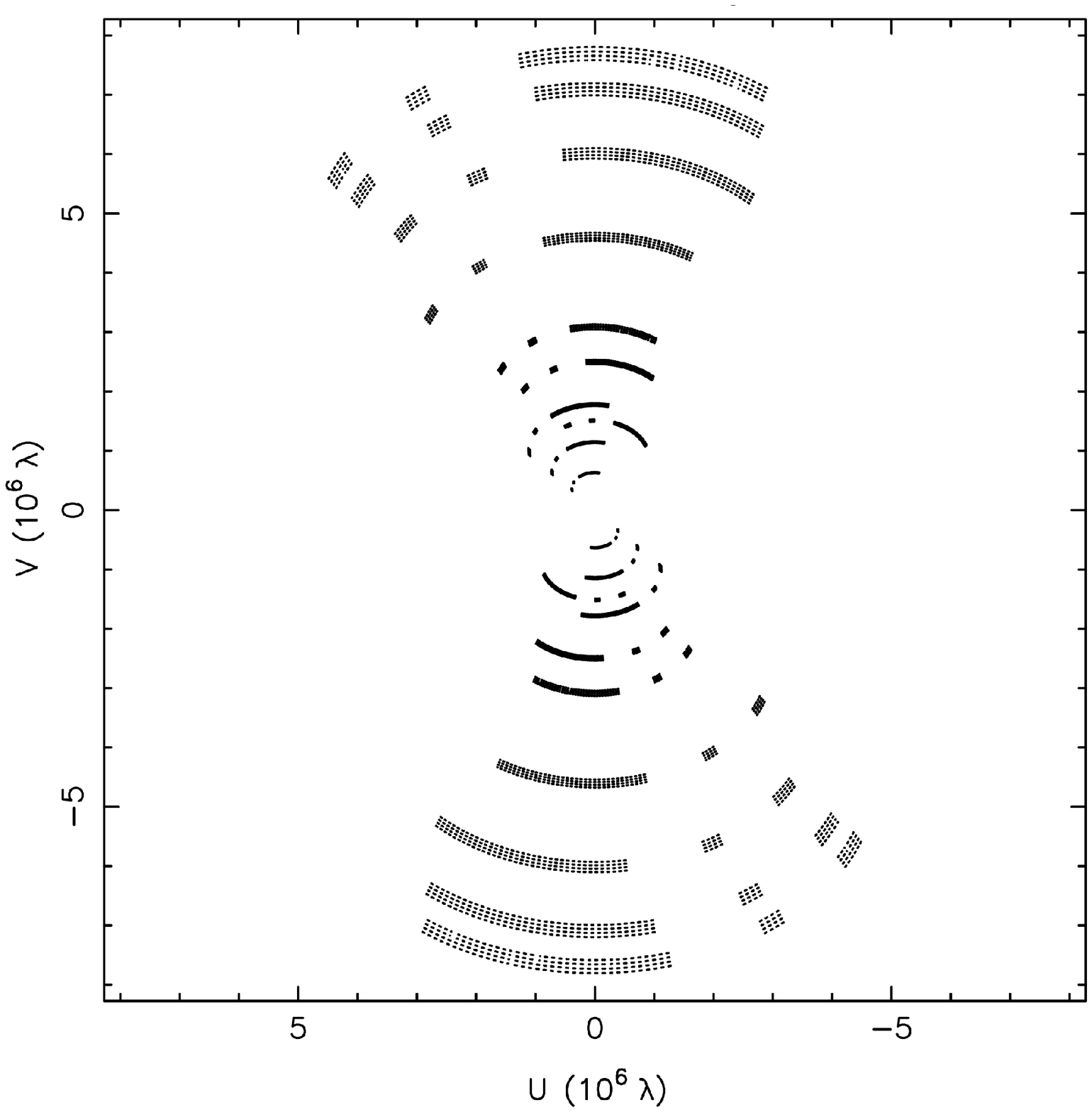} &
\includegraphics[width=0.4\textwidth]{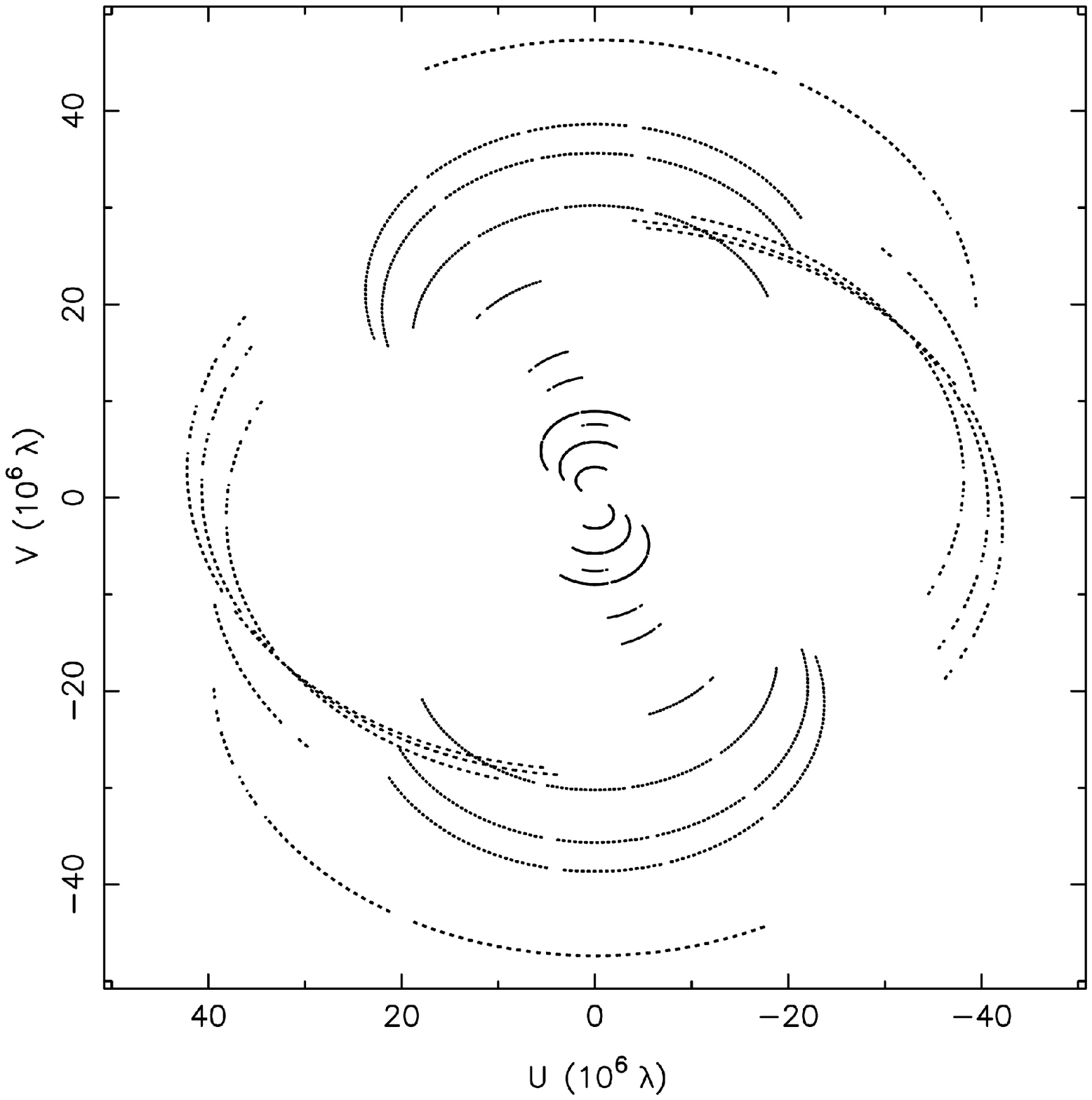} \\
\end{tabular}
\caption{Typical $uv$\ coverage at 1650 MHz (no Ceduna; left panel) and at 8400 MHz (with Ceduna;
right).  The target sources are PSR J1559--4438 and PSR J0437-4715 respectively. Without
Ceduna, the $uv$\ coverage is heavily biased North--South, and wide hour--angle coverage is 
necessary to gain acceptable $uv$\ coverage.  The uniformly weighted beam size is 
40$\times$13 mas at 1650 MHz, and 3.2$\times$2.6 mas at 8400 MHz.}
\label{fig:uvcoverage}
\end{center}
\end{figure}

\end{document}